\documentclass[a4paper,11pt]{article}

\usepackage{jheppub}

\usepackage{amsmath,amssymb}% Include figure files
\usepackage{graphicx}% Include figure files
\usepackage{dcolumn}% Align table columns on decimal point
\usepackage{bm}
\usepackage{color}

\usepackage{multirow}
\usepackage{longtable}
\usepackage{xcolor}%定义了一些颜色
\usepackage{colortbl,booktabs}%第二个包定义了几个*rule

\newcommand{\la}{\langle}
\newcommand{\ra}{\rangle}
\newcommand{\qbar}{\bar{q}}
\newcommand{\ubar}{\bar{u}}

\newcommand{\sbar}{\overline{s}}
\newcommand{\cbar}{\overline{c}}
\newcommand{\dbar}{\overline{d}}

\def\B{{\cal B}}

\def\be{\begin{eqnarray}}
\def\en{\end{eqnarray}}

\begin{document}

\title{Doubly Cabibbo-suppressed decays of  antitriplet charmed baryons}

\author{ Guanbao Meng$^{1}$, Sam Ming-Yin Wong$^{2}$, Fanrong Xu$^{1}$}
\affiliation{
 ${}^{1}$Department of Physics, Jinan University,
 Guangzhou 510632, People's Republic of China\\
  ${}^{2}$Department of Physics, National Taiwan University,
 Taipei 10617,  Republic of China
}%

\emailAdd{fanrongxu@jnu.edu.cn}

\abstract{
Doubly Cabibbo-suppressed (DCS) nonleptonic  weak decays of antitriplet charmed baryons are studied systematically in this work.  The factorizable and nonfactorizable contributions can be classified explicitly in the topological-diagram
approach and treated separately. In particular,
the evaluation of nonfactorizable terms is based on the pole model in conjunction with current algebra.
All three types of relevant non-perturbative parameters contributing factorizable 
and nonfactorizable terms are estimated in the MIT bag model. 
Branching fractions of all the DCS decays are predicted to be of order $10^{-4}\sim 10^{-6}$.
In particular, we find that the three modes $\Xi_c^+\to \Sigma^+ K^0, \Sigma^0 K^+$ and $\Xi_c^0\to \Sigma^- K^+$
are  as large as $(1\sim 2)\times 10^{-4}$, which are the most promising DCS channels to be measured.
We also point out that the two DCS modes $\Xi_c^+\to \Sigma^+ K^0$ and $\Xi_c^0\to \Sigma^0 K^0$ are possible
to be distinguished from $\Xi_c^+\to \Sigma^+ K_S$ and $\Xi_c^0\to \Sigma^0 K_S$.
The decay asymmetries for all the channels 
with a kaon in their final states are found to be large in magnitude and negative in sign.
}


\maketitle
\flushbottom
 
\section{Introduction}\label{sec:Intro}

It is known that weak decays dominate the decays of antitriplet charmed baryons, consisting of
$\Lambda_c^+, \Xi_c^0$ and $\Xi_c^+$.
Since four quarks get involved in weak decays of the antitriplet charmed baryons at tree level, we usually 
classify the weak decays into Cabibbo-favored (CF), singly Cabibbo-suppressed (SCS) and 
doubly Cabibbo-suppressed (DCS) modes according to the power of $\sin\theta_c\cos\theta_c$,
where $\theta_c$ is the Cabibbo angle. 

Recently, some progresses have been made in the experimental study of charm baryons.
First, both Belle \cite{Zupanc} and BESIII \cite{Ablikim:2015flg}  have measured
the absolute branching fraction of the decay $\Lambda_c^+\to pK^-\pi^+$, leading to
a new average of $(6.28\pm0.32)\%$ for this benchmark mode quoted by the Particle Data Group (PDG)  \cite{Tanabashi:2018oca}.
The measurement of
$\Lambda_c^+\to p\pi^0, p\eta$ \cite{Ablikim:2017ors} performed in BESIII 
indicated that SCS decays are ready to access.
Belle has also made some new developments  in the study of $\Xi_c^0$  and $\Xi_c^+$, the two other singly charmed baryons in the antitriplet.
By using a data set comprising $(772\pm11)\times 10^6$ $B\bar{B}$ pairs collected at
$\Upsilon(4S)$ resonance, Belle was able to measure 
the branching ratios of charged and neutral $\Xi_c$ decays\cite{Li:2018qak,Li:2019atu}.
In addition to the discovery of doubly charmed baryon, recently LHCb also make 
a significant contribution in singly charmed baryon. 
Three new $\Xi_c^0$ baryon states have been observed through their
decay into $\Lambda_c^+ K^-$ \cite{Aaij:2020yyt}. 
In particular, BESIII recently has published a white paper on its future prospect \cite{Ablikim:2019hff},
 indicating that the measurement of DCS decays is also anticipated. 

In recent theoretical studies,  two main approaches have been adopted. 
In one methodology, connections among various decay amplitudes can be established  
based on SU(3) flavor symmetry.
Then by taking a global fit to the existing experimental data as inputs,
more channels can be predicted \cite{Geng:2019xbo}. %It is not necessary to reveal the details of dynamics and hence
%has an advantage to avoiding the challenges in non-perturbative physics. 
Another new  attempt has also been proposed by fitting topological diagrams % as independent fitting elements
and some predictions are also given \cite{Zhao:2018mov}. 
The second approach to study charmed baryon weak decays is relying on model estimation, with which
dynamics at the quark level can be revealed. 
To understand the underlying dynamical mechanism in hadronic weak decays, one may draw the topological diagrams according to
the hadron's content \cite{Chau:1995gk}.
In charmed baryon decays, nonfactorizable contributions from $W$-exchange or inner $W$-emission diagrams play an essential role and they cannot be neglected, in contrast with the negligible effects in heavy meson decays.
To estimate the nonfactorizable effects in charmed baryon decays, 
various techniques were developed in the 1990s , including relativistic quark model (RQM) \cite{Korner:1992wi,Ivanov:1997ra}, pole model \cite{Xu:1992vc,Cheng:1993gf,Zenczykowski:1993jm} and
current algebra \cite{Cheng:1993gf,Sharma:1998rd}. 
And recently an estimation of $\Lambda_c^+$ weak decays based on nonrelativistic constitutent quark model  has been carried out \cite{Niu:2020gjw}.

Our estimation of nonfactorizable contribution will be based on the pole model.
In the pole model, important low-lying $1/2^+$
and $1/2^-$ states are usually considered under the pole approximation.
In the decay with a pseudoscalar in the final state, $\mathcal{B}_c\to\mathcal{B}'+P$, the nonfactorizable  $S$- and $P$-wave amplitudes are dominated by $1/2^-$ low-lying baryon resonances and $1/2^+$ ground state baryons, respectively.
The $S$-wave amplitude can be further reduced to current algebra in the soft-pseudoscalar limit. That is, the evaluation of the $S$-wave amplitude does not require the information of the troublesome negative-parity baryon resonances which are not well understood in the quark model.
The methodology was developed and applied in the earlier work \cite{Cheng:1993gf}. 
Our work is hence based on pole model in conjunction with current algebra.
In our previous works, we have systematically studied weak decays of antitriplet charmed baryons \cite{Cheng:2018hwl, Zou:2019kzq},
the only weak decaying baryon in sextet $\Omega_c$ \cite{Hu:2020nkg} and doubly charmed baryons \cite{Cheng:2020wmk}.
It turns out if the $S$-wave amplitude is evaluated in the pole model or in the covariant quark model and its variant, the decay asymmetries for both $\Lambda_c^+\to \Sigma^+\pi^0$ and $\Sigma^0\pi^+$ were always predicted to be positive, while it was measured to be $-0.45\pm0.31\pm0.06$ for $\Sigma^+\pi^0$ by CLEO \cite{CLEO:alpha}. In contrast, current algebra always leads to a negative decay asymmetry for aforementioned two modes: $-0.49$ in \cite{Cheng:1993gf}, $-0.31$ in \cite{Sharma:1998rd}, $-0.76$ in \cite{Zenczykowski:1993hw} and $-0.47$ in \cite{Datta}.  The issue with the sign of $\alpha(\Lambda_c^+\to\Sigma^+\pi^0)$ was finally resolved by BESIII. The decay asymmetry parameters of $\Lambda_c^+\to \Lambda\pi^+,\Sigma^0\pi^+,\Sigma^+\pi^0$ and $pK_S$ were recently measured by BESIII \cite{Ablikim:2019zwe}, for example,  $\alpha(\Lambda_c^+\to\Sigma^+\pi^0)=-0.57\pm0.12$ was obtained. 
Hence, the negative sign of $\alpha(\Lambda_c^+\to\Sigma^+\pi^0)$ measured by CLEO is nicely confirmed by BESIII.
This is one of the strong reasons why we adapt current algebra to work out parity-violating amplitudes.
For the antitriplet charmed baryon, the calculations for CF and SCS modes have been
completed \cite{Cheng:2018hwl, Zou:2019kzq}. 
In this paper, with the  prospect from experiments indicated by 
of BESIII, we will continue completing the remaining piece, DCS decays of
 antitriplet charmed baryons.

This paper is organized as follows.  In Sec. \ref{sec:Formalism} we will set up
the formalism for evaluating branching fractions and up-down decay asymmetries,
including
contributions from both
factorizable and nonfactorizable terms. 
%he factorizable contribution is calculated under naive factorization while 
%nonfactorizable contribution is estimated in pole model.
Numerical results are
presented in Sec. \ref{sec:num}.
A conclusion will be given in Sec. \ref{sec:con}.
In Appendix \ref{app:nonp}, we summarize all involved non-perturbative quantities calculated
in MIT bag model,
including baryon transition form factors, baryon matrix elements and the axial-vector form factors.

\section{Theoretical  framework}
\label{sec:Formalism}

In this section, we will first introduce the generic kinematics of two-body hadronic decays.
Then in the topological-diagram approach, factorizable and nonfactorizable  amplitudes
can be classified explicitly 
\cite{Cheng:1991sn,Cheng:1993gf}.
The further calculation of the two parts of contributions are treated separately.
For the factorizable amplitudes we  evaluate them within naive factorization, while the pole model associated
with current algebra technique is adopted in the calculation of nonfactorizable amplitudes.

\subsection{Kinematics}\label{sec:Kin}
Without loss of generality, 
the amplitude for the decay of an initial baryon $\mathcal{B}_i$  into a final baryon $\mathcal{B}_f$ and
a pseudoscalar meson $P$ can
be parametrized as
\begin{equation}
M(\mathcal{B}_i \to \mathcal{B}_f P)= i\ubar_f (A-B\gamma_5) u_i,
\end{equation}
where $A$ and $B$ stand for $S$- and $P$-wave amplitude, respectively. 
Both the two amplitudes contribute to the decay width and 
 and up-down decay asymmetry, giving
\begin{align}
& \Gamma=\frac{p_c}{8\pi}
\left[ \frac{(m_i+m_f)^2-m_P^2}{m_i^2}|A|^2
+\frac{(m_i-m_f)^2-m_P^2}{m_i^2}|B|^2\right], \nonumber \\
& \alpha=\frac{2\kappa {\rm{Re}} (A^* B)}{|A|^2+\kappa^2|B|^2},
\label{eq:Gamma}
\end{align}
where $\kappa$ is defined as $\kappa=p_c/(E_f+m_f)=\sqrt{(E_f-m_f)/(E_f+m_f)}$ and
$p_c$ is the three-momentum
in the rest frame of the mother particle.
Obviously for the magnitude of decay width the contribution from $S$-wave amplitude is larger than the $P$-wave one
up to a factor of $[(m_i+m_f)^2-m_P^2]/[(m_i-m_f)^2-m_P^2]$, while the sign of decay asymmetry
is determined by the relative sign between $A$ and $B$.

 The $S$- and $P$- wave amplitudes of the two-body decay generally receive both factorizable and nonfactorizable contributions, giving
 \begin{align}
& A=A^{\rm{fac}}+A^{\rm{nf}},\quad
B=B^{\rm{fac}}+B^{\rm{nf}}.
\label{eq:amplitude}
\end{align}
The nonfactorizable amplitudes, denoted as $A^{\rm{nf}}$ and $B^{\rm{nf}}$, play an essential role in
the decays of charmed baryon and hence cannot be ignored. This feature also differs from the
situation in bottom baryon decays.
The calculation of nonfactorizable amplitudes is a non-easy task and will be tackled in the following context.

\subsection{Topological diagrams}

%====================================================================
\begin{figure}[t]
\begin{center}
\includegraphics[width=0.95\textwidth]{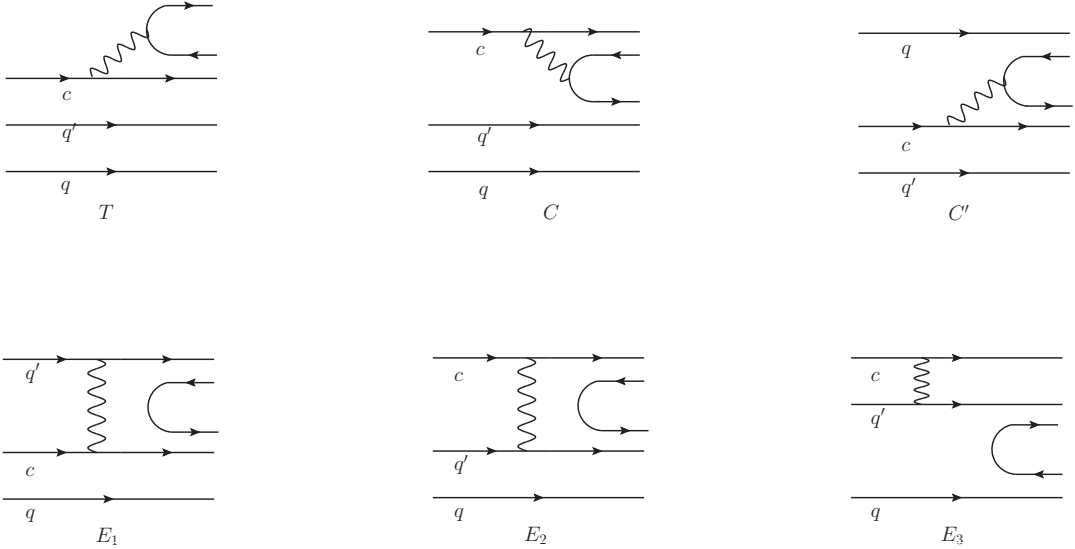}
\vspace{0.1cm}
\caption{Topological diagrams contributing to antitriplet charmed baryons decays: external $W$-emission $T$, internal $W$-emission $C$, inner $W$-emission $C'$,  $W$-exchange diagrams $E_1$, $E_2$ and $E_3$. }
 \label{fig:anti-triplet}
\end{center}
\end{figure}
%=====================================================================

The topological-diagram approach has been applied successfully in charmed meson decays. Various topological diagrams can be extracted from Cabibbo-favored (CF) channels. Then assuming SU(3) symmetry, we can use them to predict branching fractions of singly Cabibbo-suppressed (SCS) and doubly Cabibbo-suppressed (DCS) decays. Moreover, topological amplitudes allow to predict tree-induced CP violation as the information of strong phases can also be extracted. This is the power of the topological approach. 
For the charmed baryon decays, the application of the topological-diagram scheme  was proposed and systematically summarized by Chau,  Cheng and Tseng 
 more than two decades ago \cite{Chau:1995gk}. 
However,  there are not adequate data on branching fractions and decay asymmetries of 
charmed baryon decays to enable us extracting the topological diagrams. Nevertheless, we can still make use of the topological diagrams to classify the decay amplitudes into the factorizable and nonfactorizable ones.

 For the weak decays $\mathcal{B}_{c}\to \mathcal{B}+P$ ($\mathcal{B}$ is baryon octet) of interest in this work, the relevant topological diagrams are
the external $W$-emission $T$, the internal $W$-emission $C$, the inner $W$-emission $C'$,  and the $W$-exchange diagrams $E_1$, $E_2$ and $E_3$ as depicted in Fig. \ref{fig:anti-triplet}. Among them, $T$ and $C$ are factorizable, while $C'$ and $W$-exchange diagrams give nonfactorizable contributions.  The relevant topological diagrams  for DCS  decay modes  of
antitriplet charm baryons
are shown  in Table \ref{tab:modes}.

From Table \ref{tab:modes} we notice that (i) among all the DCS decays of antitriplet charmed baryons, there is no  purely factorizable mode,
(ii) the decays containing pion or $\eta$ in their final states  receive purely nonfactorizable contributions, and (iii)  the modes containing kaon receive both factorizable and nonfactorizable contributions.

%==========================================================
\begin{table}[t]
\caption{Topological diagrams contributing to DCS modes of
two-body weak decays
 $\mathcal{B}_c\to \mathcal{B} P$, where $\mathcal{B}$ is a baryon octet
and $P$ is a pseudoscalar meson.
\label{tab:modes}}
\vspace{0.3cm}
%\begin{ruledtabular}
\begin{tabular}{l l l l l l}
%\hline\hline
\toprule
$\Xi_c^+$ &  Contributions  & $\Xi_c^0$ & Contributions & $\Lambda_c^+$ & Contributions \\
%\colrule
%\hline
\midrule
 $\Xi_{c}^{+}\to p \eta\quad $&  $C^{'}, E_1,E_{2}, E_3$ &$\Xi_{c}^{0}\to \Sigma^{-} K^{+}$ &  $T, E_1$ &$\Lambda_{c}^{+}\to p K^{0}$ &  $C,C^{'}$\\
 $\Xi_{c}^{+}\to p \pi^{0}$ & $ E_1,E_{2}, E_3$   &$\Xi_{c}^{0}\to n \eta$ 
 &  $  C^{'}, E_1,E_{2}, E_3$  &$\Lambda_{c}^{+}\to n K^{+}$ &  $ T, C^{'}$\\
 $\Xi_{c}^{+}\to n \pi^{+}$&$E_1, E_3$  &$\Xi_{c}^{0}\to n \pi^{0}$&$ E_1,E_{2}, E_3$ &$$ \\
  $\Xi_{c}^{+}\to \Sigma^{0} K^{+}$ & $ T,  C^{'}, E_1, E_3$ &$\Xi_{c}^{0}\to \Sigma^{0} K^{0}$ & $C,C^{'},E_{2}, E_3$ &$$&$$\\
 $\Xi_{c}^{+}\to \Lambda^{0} K^{+}$&$ T, C^{'}, E_1, E_3 $ &$\Xi_{c}^{0}\to \Lambda^{0} K^{0}$ & $C,C^{'},E_{2}, E_3$ &$$ &$$\\
$\Xi_{c}^{+}\to \Sigma^{+} K^{0}$& $ C,E_{2}$& $\Xi_{c}^{0}\to p \pi^{-}$& $ E_{2}, E_3$  &$$&$$ \\
%\hline\hline
\bottomrule
\end{tabular}
%\end{ruledtabular}
\end{table}
%=================================================

\subsection{Factorizable amplitudes}

In the frame of topological diagrams, the external $W$-emission $T$ and internal $W$-emission $C$ represent factorizable contributions.
 Strictly speaking, there are also nonfactorizable effects in the two diagrams.
However, these nonfactorizable effects  can be absorbed by  an effective $N_c$ in the effective Wilson coefficients, 
and the value of $N_c$ can be extracted from the data. In that sense, the form of naive factorization can be kept  and hence $T$ and $C$ can be classified into factorizable ones.

The effective Hamiltonian to describe the DCS decays of antitriplet charmed baryons is
\begin{equation}
\mathcal{H}_{\rm{eff}}=\frac{G_F}{\sqrt{2}}V_{cd}V_{us}^*(c_1O_1+c_2O_2)+H.c.,
\label{eq:Hamiltonian}
\end{equation}
where the four-quark operators are given by
\begin{equation}
O_1=(\ubar s)(\dbar c),\quad
O_2=(\ubar c)(\dbar s),
\end{equation}
while the abbreviated notation in four-quark operators is defined as
 $ (\qbar_1 q_2)\equiv \qbar_1\gamma_\mu(1-\gamma_5) q_2\nonumber$.
The
Wilson coefficients to the leading order are given as $c_1=1.346$ and $c_2=-0.636$ at $\mu=1.25\,\rm{GeV}$ and
$\Lambda_{\rm{MS}}^{(4)}=325\,{\rm{MeV}}$ \cite{Buchalla:1995vs}.
Considering the mixing of operators, it is more convenient to introduce effective Wilson coefficients 
$a_1=c_1+\frac{c_2}{N_c}$ and $a_2=c_2+\frac{c_1}{N_c}$ where $N_c$ is the number of colors.
Topological diagrams contain all the final state interactions and in principle should also include non-factorizable contributions.
However, it turns out in charm physics such effect is small.
In order to incorporate the small non-factorizable effects we furthermore
define an effective $N_c$ and its value can be extracted from the experimental data.
A recent measurement of $\mathcal{B}(\Lambda_c\to p \phi)=(1.04\pm 0.21)\times 10^{-3}$ by BESIII \cite{Ablikim:2017ors},
which receives purely factorizable contribution,
indicates $N_c^{\rm{eff}}\approx 7$, and hence we have $a_1=1.26$ and $a_2=-0.45$ \cite{Cheng:2018hwl}.

Now under naive factorization
the amplitude can be written down as
\begin{equation}
M=\la P\mathcal{B}|\mathcal{H}_{\rm{eff}}|\mathcal{B}_{c}\ra
=\left\{\begin{array}{ll}
\frac{G_F}{\sqrt{2}}V_{cd}V_{us}^* a_{1} \la P|(\ubar s)|0\ra \la \mathcal{B}|(\dbar c)|\mathcal{B}_{c}\ra , &P =K^+,\\
\\
\frac{G_F}{\sqrt{2}}V_{cd}V_{us}^* a_{2} \la P|(\sbar d)|0\ra \la \mathcal{B}|(\ubar c)|\mathcal{B}_{c}\ra, &P={K}^0, \\
\end{array}
\label{eq:DCS}
\right.
\end{equation}
where $a_1$ corresponds to charged kaon while $a_2$ characterizes the amplitude with neutral kaon final state. 
In terms of the decay constants
\begin{equation}
\la K (q)|\sbar\gamma_\mu(1-\gamma_5) d|0\ra = if_K q_\mu
\label{KFF}
\end{equation}
and the form factors defined by
\begin{eqnarray}
\la \mathcal{B}(p_2)|\cbar\gamma_\mu(1-\gamma_5) u|\mathcal{B}_{c}(p_1)\ra
&=&\ubar_2 \left[ f_1(q^2) \gamma_\mu -f_2(q^2)i\sigma_{\mu\nu}\frac{q^\nu}{M}+f_3(q^2)\frac{q_\mu}{M}\right.\\
&&\hspace{0.5cm} -\left.\left(g_1(q^2)\gamma_\mu-g_2 (q^2)i\sigma_{\mu\nu}\frac{q^\nu}{M}+g_3(q^2)
\frac{q_\mu}{M}
\right)\gamma_5
\right]u_1,  \nonumber
\end{eqnarray}
with
%the  initial particle mass  $M$ and
the momentum transfer  $q=p_1-p_2$,
we obtain the  amplitude
\begin{equation}
M(\mathcal{B}_{c}\to \mathcal{B} P)
=i\frac{G_F}{\sqrt{2}}a_{1,2} V_{us}^*V_{cd} f_P \ubar_2(p_2)\left[(m_1-m_2)f_1(q^2)
+(m_1+m_2)g_1(q^2) \gamma_5\right]u_1(p_1).
\end{equation}
The contributions from the form factors $f_{3}$ and $g_{3}$ can be neglected 
for the similar reasons in the case of CF and CSC decays \cite{Zou:2019kzq}.
Hence the factorizable contributions to $A$ and $B$ terms finally read
\begin{align}
&A^{\rm{fac}}=\frac{G_F}{\sqrt{2}}a_{1,2} %\sum_{q=d,s}
V_{us}^*V_{cd} f_P(m_{\mathcal{B}_{c}}-m_{\mathcal{B}}) f_1(q^2),
\nonumber\\
&B^{\rm{fac}}= -\frac{G_F}{\sqrt{2}}a_{1,2} %\sum_{q=d,s}
V_{us}^*V_{cd} f_P(m_{\mathcal{B}_{c}}+m_{\mathcal{B}})
g_1(q^2).
\end{align}
The factorizable amplitudes only appear in the modes containing kaon, and 
 the choice of $a_i$ is determined by the electric charge of 
final states kaon, see Eq. (\ref{eq:DCS}).

The size of
 the factorizable amplitudes, together with the nonfactorizable ones, determines the branching 
 fractions and decay asymmetries.
Meanwhile its sign also plays a crucial role, which tells whether the interference with non-factorizable ones 
is destructive or constructive.
In this work, the evaluation of baryon transition form factors $f_1$ and $g_1$ is
carried out within the MIT bag model in the static limit. 
The exact calculated results for form factors are summarized 
in Appendix \ref{app:FF}, where 
we first show the detailed results in the zero recoil limit $q^2=(m_i-m_f)^2$ and then a further correction to $q^2=m_P^2$ 
is made. 

\subsection{Nonfactorizable amplitudes}
\label{subsec:nf}

%====================================================================
\begin{figure}[t]
\begin{center}
\includegraphics[width=0.90\textwidth]{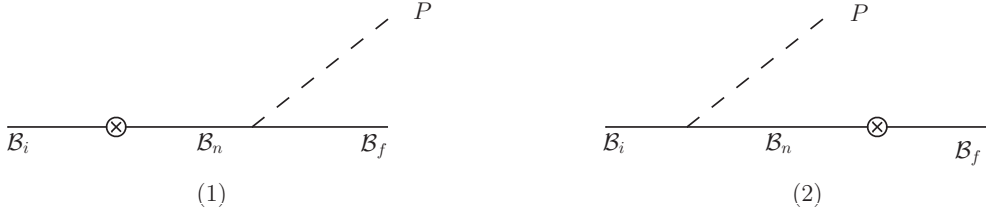}
\vspace{0.1cm}
\caption{Pole diagrams for two-body charmed baryon hadronic decays with an initial baryon $\mathcal{B}_i$,
a final baryon $\mathcal{B}_f$ and a final pseudoscalar meson $P$. The cross inserted in the straight line
stands for weak interaction.}
 \label{fig:pole}
\end{center}
\end{figure}
%=====================================================================

Nonfactorizable amplitudes give critical contributions in charmed baryon decays.
In the topological-diagram approach, the three types of 
diagrmas $C', E_1, E_2$\footnote{
The contribution from $E_3$ will be discussed hereafter.
}
are classified to depict nonfactorizable contributions. 
Various methods have been developed to study nonfactorizable contribution, here we
keep on working in the pole model.
There are two kinds of pole diagrmas  in the pole model approximation.
A correspondence is made between topological and pole diagrams, in which 
$C'$ maps to type (2) in Fig. \ref{fig:pole} while $E_1$ and $E_2$ identically corresponds to type (1)
  and $E_3$  receives both pole contributions. 

In the pole model, the general formula for $S$- and $P$-wave amplitudes 
can be extracted from a complete amplitude according to Fig. \ref{fig:pole},
%\cite{Cheng:2018hwl,Zou:2019kzq,Cheng:2020wmk,Hu:2020nkg}
\begin{align}
& A^{\rm{pole}}=-\sum\limits_{\mathcal{B}_n^*(1/2^-)}\left[\frac{g_{\mathcal{B}_f \mathcal{B}_n^* M}b_{n^* i}}{m_i-m_{n^*}} +
\frac{b_{fn^*}g_{\mathcal{B}_{n}^*\mathcal{B}_i M}}{m_f-m_{n^*}}\right],\nonumber\\
& B^{\rm{pole}}=\sum\limits_{\mathcal{B}_n}\left[ \frac{g_{\mathcal{B}_f \mathcal{B}_n M}a_{ni}}{m_i-m_n}
+\frac{a_{fn}g_{\mathcal{B}_n \mathcal{B}_i M}}{m_f-m_n}\right],
\end{align}
where $g_{ij n}$ is the strong coupling among the pseudoscalar meson and two baryons, and  the baryonic matrix elements $a_{ij}$ and $b_{ij}$ are defined as
\begin{equation}
\la \mathcal{B}_n|H|\mathcal{B}_i\ra = \ubar_n (a_{ni} + b_{n i}\gamma_5) u_i,\qquad
\la \mathcal{B}_i^*(1/2^-) | H|\mathcal{B}_j\ra =\ubar_{i*} b_{i^* j} u_j. \label{eq:bmatrix}
\end{equation}
To estimate  the $S$-wave amplitudes in the pole model is a difficult and nontrivial task as it involves the matrix elements and strong coupling constants of $1/2^-$ baryon resonances which is less known \cite{Cheng:1991sn}.
Nevertheless, provided a soft emitted pseudoscalar meson, the intermediate excited baryons can be summed up, leading to a commutator term
\be
A^{\rm{com}} &=& -\frac{\sqrt{2}}{f_{P^a}}\la \mathcal{B}_f|[Q_5^a, H_{\rm{eff}}^{\rm PV}]|\mathcal{B}_i\ra
=\frac{\sqrt{2}}{f_{P^a}}\la \mathcal{B}_f|[Q^a, H_{\rm{eff}}^{\rm PC}]|\mathcal{B}_i\ra, \label{eq:Apole}
\en
with the conserving charges
\begin{equation}
Q^a=\int d^3x \qbar\gamma^0\frac{\lambda^a}{2}q,\qquad
Q^a_5=\int d^3x \qbar\gamma^0\gamma_5\frac{\lambda^a}{2}q.
\end{equation}
Likewise, the $P$-wave amplitude is reduced in the soft-meson limit to
\be
B^{\rm{ca}} &=& \frac{\sqrt{2}}{f_{P^a}}\sum_{\mathcal{B}_n}\left[ g^A_{\mathcal{B}_f \mathcal{B}_n}\frac{m_f+m_n}{m_i-m_n}a_{ni}
+a_{fn}\frac{m_i + m_n}{m_f-m_n} g_{\mathcal{B}_n \mathcal{B}_i}^A\right],
\label{eq:Bpole}
\en
where  the generalized Goldberger-Treiman relation,
\begin{equation} \label{eq:GT}
g_{_{\mathcal{B'B}P^a}}=\frac{\sqrt{2}}{f_{P^a}}(m_{\mathcal{B}}+m_{\mathcal{B'}})g^A_{\mathcal{B'B}},
\end{equation}
has been applied.
Our followup calculations will be based on
Eqs. (\ref{eq:Apole}) and (\ref{eq:Bpole}) 
%are the master equations for nonfactorizable amplitudes 
in the pole model under the soft meson approximation.

\subsubsection{$S$-wave amplitudes}
As shown in Eq. (\ref{eq:Apole}),
the nonfactorizable $S$-wave amplitude can be simplified into the commutator terms of conserving charge $Q^a$ and the parity-conserving part of the Hamiltonian.
Under SU(3) symmetry, the involved conserving charges in different decays are determined 
by final state mesons.  In terms of commutators,
the exact expressions for
various $S$-wave amplitudes are:
\begin{align} \label{eq:commu}
&A^{\rm{com}}(B_i\to B_f \pi^{\pm})=\frac{1}{f_\pi}\la B_f|[I_{\mp}, H_{\rm{eff}}^{PC}]|B_i\ra,\nonumber\\
&A^{\rm{com}}(B_i\to B_f \pi^{0})=\frac{\sqrt{2}}{f_\pi}\la B_f|[I_3, H_{\rm{eff}}^{PC}]|B_i\ra,\nonumber\\
&A^{\rm{com}}(B_i\to B_f \eta_8)=\sqrt{\frac32}\frac{1}{f_{\eta_8}}\la B_f|[Y, H_{\rm{eff}}^{PC}]|B_i\ra,\nonumber\\
&A^{\rm{com}}(B_i\to B_f K^{\pm})=\frac{1}{f_K}\la B_f|[V_{\mp}, H_{\rm{eff}}^{PC}]|B_i\ra,\nonumber\\
&A^{\rm{com}}(B_i\to B_f \overline{K}^0 )=\frac{1}{f_K}\la B_f|[U_{+}, H_{\rm{eff}}^{PC}]|B_i\ra,\nonumber\\
&A^{\rm{com}}(B_i\to B_f {K^0})=\frac{1}{f_K}\la B_f|[U_{-}, H_{\rm{eff}}^{PC}]|B_i\ra.
\end{align}
In particular, the octet component can be extracted from the mixing in $\eta$ and $\eta'$
\be
\eta=\cos\theta\eta_8-\sin\theta\eta_0, \qquad \eta'=\sin\theta\eta_8+\cos\theta\eta_0,
\en
with $\theta=-15.4^\circ$ \cite{Kroll}. For the decay constant $f_{\eta_8}$,  we shall follow \cite{Kroll} to use $f_{\eta_8}=f_8\cos\theta$ with $f_8=1.26 f_\pi$.
As its conserving charge, hypercharge $Y$, we shall follow the convention  $Y=B+S-C$ \cite{Cheng:2018hwl}.

The calculation of commutators requires the information of baryon's behaviors under $U, V, I$ symmetries.
In this work, we still use the wave function conventions in our previous works  \cite{Cheng:2018hwl,Zou:2019kzq,Cheng:2020wmk,Hu:2020nkg}, and especially their features under ladder operators can be found in Appendix B of \cite{Zou:2019kzq}.
 After a straightforward calculation of commutators,
  we obtain 
 the S-wave amplitudes, 
\begin{align}
& A^{\rm{com}}(\Xi_{c}^{+}\to p \eta)=0,\qquad\qquad  \qquad\qquad\qquad
A^{\rm{com}}(\Xi_{c}^{+}\to p \pi^{0})= 0,  \nonumber\\
&A^{\rm{com}}(\Xi_{c}^{+}\to n \pi^{+})=
\frac{1}{f_{\pi}}  a_{p \Xi_{c}^{+}} ,\qquad\qquad\qquad\;
A^{\rm{com}}(\Xi_{c}^{+}\to \Sigma^{0} K^{+})= -\frac{\sqrt{2}}{2f_{K}}a_{p \Xi_{c}^{+}} ,    \nonumber\\
&A^{\rm{com}}(\Xi_{c}^{+}\to \Lambda^{0} K^{+})= -\frac{\sqrt{6}}{2f_{K}}a_{p \Xi_{c}^{+}} ,  \qquad\qquad
A^{\rm{com}}(\Xi_{c}^{+}\to \Sigma^{+} K^{0})= -\frac{1}{f_{K}}a_{p \Xi_{c}^{+}},     \nonumber\\
&A^{\rm{com}}(\Xi_{c}^{0}\to n \pi^{0})= 0,  \qquad\qquad \qquad  \quad\qquad  \;\,
 A^{\rm{com}}(\Xi_{c}^{0}\to n \eta_{8})= 0,\nonumber\\
&A^{\rm{com}}(\Xi_{c}^{0}\to \Sigma^{-} K^{+})= -\frac{1}{f_{K}}a_{n \Xi_{c}^{0}} ,  \qquad\qquad\;
A^{\rm{com}}(\Xi_{c}^{0}\to\Sigma^{0} K^{0})= \frac{\sqrt{2}}{2f_{K}}a_{n \Xi_{c}^{0}} ,    \nonumber\\
&A^{\rm{com}}(\Xi_{c}^{0}\to\Lambda^{0} K^{0})=- \frac{\sqrt{6}}{2f_{K}}a_{n \Xi_{c}^{0}}   ,  \qquad\qquad\;
A^{\rm{com}}(\Xi_{c}^{0}\to p \pi^{-})= \frac{1}{f_{\pi}}a_{n \Xi_{c}^{0}},    \nonumber\\
% \frac{1}{f_{\pi}}(a_{n \Xi_{c}^{0}}-a_{p \Xi_{c}^{+}}),    \nonumber\\
&A^{\rm{com}}(\Lambda_{c}^{+}\to p K^{0})= \frac{1}{f_{K}}a_{p \Xi_{c}^{+}} ,  \qquad\qquad\qquad
A^{\rm{com}}(\Lambda_{c}^{+}\to n K^{+})= -\frac{1}{f_{K}}a_{n \Xi_{c}^{0}},
\label{eq:com}
\end{align}
in which the quantities $a_{\mathcal{B}'\mathcal{B}}$ is defined in Eq. (\ref{eq:bmatrix}).
In particular, 
the vanishing $S$-wave amplitudes of the four modes decaying into $p\pi^0, p\eta, n\pi^0$ and $n\eta$
are due to the identical quantum numbers of initial and final baryons, specifically given as
$ I_3(n)=I_3(\Xi_c^0)=-\frac12, %\quad 
I_3(p)=I_3(\Xi_c^+)=\frac12$ for $\pi^0$
and
$Y(n)=Y(\Xi_c^0)=1, %\qquad 
Y(p)=Y(\Xi_c^+)=1$ for $\eta$.
This  natural consequence of current algebra, however, is too strong and 
a minor correction of current algebra will change the prediction dramatically, especially for the decay asymmetries.
 As for the two modes $\Xi_c^+\to n\pi^+$ and $\Xi_c^0 \to p\pi^-$, 
 a straightforward current algebra calculation yields two terms which cancel each other. 
 As explained in the beginning of Sec. \ref{subsec:nf}, they correspond to $E_3$ which can be neglected. In Eq. (\ref{eq:com}) current algebra results arise from $E_1$ for $\Xi_c^+\to n \pi^+$ and $E_2$ for $\Xi_c^0\to p \pi^-$. 
 %
% we have already dropped one term in each channel, which will induce a cancellation in %$A^{\rm{com}}$,
% for the neglection of topological diagram $E_3$. 
 %
 We will further illustrate the mechanism in the 
 following section.
 A further estimation of baryon matrix elements $a_{\mathcal{B}'\mathcal{B}}$ in MIT bag model is carried out in Appendix \ref{app:bme}.

\subsubsection{$P$-wave amplitudes}
According to Eq. (\ref{eq:Bpole}),
the nonfactorizable $P$-wave amplitudes can be obtained by considering various intermediate states,
\begin{align}
&B^{\rm{ca}}(\Xi_{c}^{+}\to p \eta)=\frac{\sqrt{2}}{f_{\eta_{8}}}\left(a_{p \Xi^{+}_{c}}\frac{m_{\Xi_{c}^{+}}+m_{\Xi_{c}^{+}}}{m_{p}-m_{\Xi_{c}^{+}}}
g^{A(\eta_{8})}_{\Xi_{c}^{+} \Xi_{c}^{+}}+a_{p \Xi^{'+}_{c}}\frac{m_{\Xi_{c}^{+}}+m_{\Xi_{c}^{'+}}}{m_{p}-m_{\Xi_{c}^{'+}}}
g^{A(\eta_{8})}_{\Xi_{c}^{'+} \Xi_{c}^{+}}\right.\nonumber\\
&\hspace{4cm}\left.
+g^{A(\eta_{8})}_{pp} \frac{m_{p}+m_{p}}{m_{\Xi_{c}^{+}}-m_{p}}a_{p \Xi^{+}_{c}}\right), \nonumber\\
&B^{\rm{ca}}(\Xi_{c}^{+}\to p \pi^{0})=\frac{\sqrt{2}}{f_{\pi}}\left(g^{A(\pi^{0})}_{pp}\frac{m_{p}+m_{p}}{m_{\Xi_{c}^{+}}-m_{p}}
a_{p \Xi^{+}_{c}} \right), \nonumber\\
&B^{\rm{ca}}(\Xi_{c}^{+}\to n \pi^{+})=\frac{1}{f_{\pi}}\left(g^{A(\pi^{+})}_{np}\frac{m_{n}+m_{p}}{m_{\Xi_{c}^{+}}-m_{p}}
a_{p \Xi^{+}_{c}}\right),\nonumber\\
&B^{\rm{ca}}(\Xi_{c}^{+}\to \Sigma^{0} K^{+})=\frac{1}{f_{K}}\left(g^{A(K^{+})}_{\Sigma^{0} p}\frac{m_{\Sigma^{0}}+m_{p}}{m_{\Xi_{c}^{+}}-m_{p}}
a_{p \Xi^{+}_{c}}+a_{\Sigma^{0} \Omega^{0}_{c}}\frac{m_{\Xi_{c}^{+}}+m_{\Omega^{0}_{c}}}{m_{\Sigma ^{0}}-m_{\Omega^{0}_{c}}}
g^{A(K^{+})}_{\Omega^{0}_{c} \Xi_{c}^{+}}\right),\nonumber\\
&B^{\rm{ca}}(\Xi_{c}^{+}\to \Lambda^{0} K^{+})=\frac{1}{f_{K}}\left(g^{A(K^{+})}_{\Lambda^{0} p}\frac{m_{\Lambda^{0}}+m_{p}}{m_{\Xi_{c}^{+}}-m_{p}}
a_{p \Xi^{+}_{c}}+a_{\Lambda^{0}\Omega^{0}_{c}}\frac{m_{\Xi_{c}^{+}}+m_{\Omega^{0}_{c}}}{m_{\Lambda^{0}}-m_{\Omega^{0}_{c}}}
g^{A(K^{+})}_{\Omega^{0}_{c} \Xi_{c}^{+}}\right),\nonumber\\
&B^{\rm{ca}}(\Xi_{c}^{+}\to \Sigma^{+} K^{0})=\frac{1}{f_{K}}\left(g^{A(K^{0})}_{\Sigma^{+} p}\frac{m_{\Sigma^{+}}+m_{p}}{m_{\Xi_{c}^{+}}-m_{p}}
a_{P \Xi^{+}_{c}}\right),\nonumber\\
&B^{\rm{ca}}(\Xi_{c}^{0}\to \Sigma^{-} K^{+})=\frac{1}{f_{K}}\left(g^{A(K^{+})}_{\Sigma^{-} n}\frac{m_{\Sigma^{-}}+m_{n}}{m_{\Xi_{c}^{0}}-m_{n}}
a_{n \Xi^{0}_{c}}\right),\nonumber\\ 
&B^{\rm{ca}}(\Xi_{c}^{0}\to n \eta)=\frac{\sqrt{2}}{f_{\eta_{8}}}\left(a_{n \Xi^{0}_{c}}\frac{m_{\Xi_{c}^{0}}+m_{\Xi_{c}^{0}}}{m_{n}-m_{\Xi_{c}^{0}}}
g^{A(\eta_{8})}_{\Xi_{c}^{0} \Xi_{c}^{0}}+a_{n \Xi^{'0}_{c}}\frac{m_{\Xi_{c}^{0}}+m_{\Xi_{c}^{'0}}}{m_{n}-m_{\Xi_{c}^{'0}}}
g^{A(\eta_{8})}_{\Xi_{c}^{'0} \Xi_{c}^{0}}\right.\nonumber\\
&\left.\hspace{4cm}
+g^{A(\eta_{8})}_{n n}\frac{m_{n}+m_{n}}{m_{\Xi_{c}^{0}}-m_{n}}
a_{n \Xi^{0}_{c}}\right),\nonumber\\ \nonumber
&B^{\rm{ca}}(\Xi_{c}^{0}\to n \pi^{0})=\frac{\sqrt{2}}{f_{\pi}}\left(g^{A(\pi^{0})}_{n n}\frac{m_{n}+m_{n}}{m_{\Xi_{c}^{0}}-m_{n}}
a_{n \Xi^{0}_{c}}\right), \nonumber\\ \nonumber
&B^{\rm{ca}}(\Xi_{c}^{0}\to \Sigma^{0} K^{0})=\frac{1}{f_{K}}\left(g^{A(K^{0})}_{\Sigma^{0} n}\frac{m_{\Sigma^{0}}+m_{n}}{m_{\Xi_{c}^{0}}-m_{n}}
a_{n \Xi^{0}_{c}}+a_{\Sigma^{0} \Omega^{0}_{c}}\frac{m_{\Xi_{c}^{0}}+m_{\Omega^{0}_{c}}}{m_{\Sigma^{0}}-m_{\Omega^{0}_{c}}}
g^{A(K^{0})}_{\Omega^{0}_{c} \Xi_{c}^{0}}\right),\nonumber\\
&B^{\rm{ca}}(\Xi_{c}^{0}\to \Lambda^{0} K^{0})=\frac{1}{f_{K}}\left(g^{A(K^{0})}_{\Lambda^{0} n}\frac{m_{\Lambda^{0}}+m_{n}}{m_{\Xi_{c}^{0}}-m_{n}}
a_{n \Xi^{0}_{c}}+a_{\Lambda^{0} \Omega^{0}_{c}}\frac{m_{\Xi_{c}^{0}}+m_{\Omega^{0}_{c}}}{m_{\Lambda^{0}}-m_{\Omega^{0}_{c}}}
g^{A(K^{0})}_{\Omega^{0}_{c} \Xi_{c}^{0}}\right),\nonumber\\
&B^{\rm{ca}}(\Xi_{c}^{0}\to p \pi^{-})=\frac{1}{f_{\pi}}\left(g^{A(\pi^{-})}_{p n}\frac{m_{p}+m_{n}}{m_{\Xi_{c}^{0}}-m_{n}}
a_{n \Xi^{0}_{c}}\right),\nonumber\\
&B^{\rm{ca}}(\Lambda_{c}^{+}\to p K^{0})=\frac{1}{f_{K}}\left(a_{p \Xi^{+}_{c}}\frac{m_{\Lambda_{c}^{+}}+m_{\Xi_{c}^{+}}}{m_{p}-m_{\Xi_{c}^{+}}}
g^{A(K^{0})}_{\Xi_{c}^{+} \Lambda_{c}^{+}}+a_{p \Xi^{'+}_{c}}\frac{m_{\Lambda_{c}^{+}}+m_{\Xi_{c}^{'+}}}{m_{p}-m_{\Xi_{c}^{'+}}}
g^{A(K^{0})}_{\Xi_{c}^{'+} \Lambda_{c}^{+}}\right),\nonumber\\
&B^{\rm{ca}}(\Lambda_{c}^{+}\to n K^{+})=\frac{1}{f_{K}}\left(a_{n \Xi^{0}_{c}}\frac{m_{\Lambda_{c}^{+}}+m_{\Xi_{c}^{0}}}{m_{n}-m_{\Xi_{c}^{0}}}
g^{A(K^{+})}_{\Xi_{c}^{0} \Lambda_{c}^{+}}+a_{n \Xi^{'0}_{c}}\frac{m_{\Lambda_{c}^{+}}+m_{\Xi_{c}^{'0}}}{m_{n}-m_{\Xi_{c}^{'0}}}
g^{A(K^{+})}_{\Xi_{c}^{'0} \Lambda_{c}^{+}}\right).
\end{align}
In general, the two types of non-perturbative parameters,
 $a_{\mathcal{B}\mathcal{B}'}$ and $g_{\B \B'}^{A(P)}$, can be calculated by Lattice QCD, QCD sum rule or quark models.
%  In this work, we will work in
  The MIT bag model, as aforementioned, is taken %to calculate these parameters 
  in this work and
the estimated results are shown explicitly
   in Appendices \ref{app:bme} and \ref{app:avff}.

\section{Results and discussion}
\label{sec:num}

In this section, we first present the details of numerical results of 
relevant branching fractions and decay asymmetries. 
%Some channels related to neutral kaon cannot directly connect to 
%experimental measurement, for this sense, we make a clarification
%in the followup parts. 
Specifically, we clarify the relation between modes with $K_S$ final state, which are more concerned by experimentalists,
and CF modes containing $\overline{K}^0$ as well as DCS modes with $K^0$.
A comparison of theoretical predictions 
with other groups is also made in the end.

\subsection{Numerical results}

%================================
\begin{table}[t]
%\footnotesize{
 \caption{
  Amplitudes (in units of  $10^{-2}G_F  {\rm{GeV}}^2$), branching fractions (in units of  $10^{-4}$) and decay asymmetries $\alpha$ of DCS modes of weak decays $\mathcal{B}_{c}\to \mathcal{B}_f P$.  
 } \label{tab:result1}
 \vspace{0.3cm}
%\centering
%\begin{ruledtabular}
%\begin{center}
\begin{tabular}%{\textwidth}
{l c c c c c c l c c }
 \toprule
% \hline\hline
 Modes & $A^{\rm{fac}}$ &  $A^{\rm{com}}$ & $A^{\rm{tot}}$ & $B^{\rm{fac}}$ &  $B^{\rm{ca}}$ & $B^{\rm{tot}}$ & $\mathcal{B}_{\rm{theo}}$
 & $\mathcal{B}_{\rm{expt}}$ &  $\alpha_{\rm{theo}}$ \\
   \midrule
  % \hline
%$\Omega_{c}^{0}\to\Omega^- \pi^+$ &  &  &  &  &  &   &    &   &
%\\
$\Xi_{c}^{+}\to p \eta$ & $ 0$  & $0$ & $0$ & $0$ & $-0.51$ & $-0.51$ & $0.32$  & $-$  & $0$
\\
$\Xi_{c}^{+}\to p \pi^{0}$ & $ 0$  & $0$ & $0$ & $0$ & $0.28$ & $0.28$ & $0.12$  & $-$  & $0$
\\
$\Xi_{c}^{+}\to n \pi^{+}$ & $ 0$  & $0.29$ & $0.29$ & $0$ & $0.39$ & $0.39$ & $0.88$  & $-$  & $0.88$
\\
$\Xi_{c}^{+}\to\Sigma^{+}K^{0}$ & $-0.14$  & $-0.24$ & $-0.39$ & $0.50$ & $0.08$ & $0.58$ & $1.28$  & $-$  & $-0.79$
\\
$\Xi_{c}^{+}\to\Sigma^{0}K^{+}$ & $ -0.28$  & $-0.17$ & $-0.45$ & $1.00$ & $0.06$ & $1.05$ & $2.26$  & $-$  & $-0.96$
\\
$\Xi_{c}^{+}\to\Lambda^{0}K^{+}$ & $ 0.15$  & $-0.30$ & $-0.15$ & $-0.51$ & $0.64$ & $0.13$ & $0.18$  & $-$  & $-0.54$
\\
$\Xi_{c}^{0}\to p \pi^{-}$ & $0$  & $0.29$ & $0.29$ & $0$ & $0.39$ & $0.39$ & $0.30$  & $-$  & $0.88$
\\
$\Xi_{c}^{0}\to n \pi^{0}$ & $0$  & $0$ & $0$ & $0$  & $-0.28$ & $-0.28$  & $0.04$ &$-$  & $0$
\\
$\Xi_{c}^{0}\to n \eta$ & $0$  & $0$ & $0$ & $0$ & $-0.52$ & $-0.52$ & $0.11$  & $-$  & $0$
\\
$\Xi_{c}^{0}\to\Sigma^{-}K^{+}$ & $-0.40$  & $-0.24$ & $-0.64$ & $1.42$ & $0.08$ & $1.50$ & $1.52$  & $-$  & $-0.96$
\\
$\Xi_{c}^{0}\to\Sigma^{0}K^{0}$ & $0.10$  & $0.17$ & $0.27$ & $-0.35$ & $-0.06$ & $-0.41$ & $0.22$  & $-$  & $-0.79$
\\
$\Xi_{c}^{0}\to\Lambda^{0}K^{0}$ & $0.05$  & $-0.30$ & $-0.25$ & $-0.18$ & $0.64$ & $0.46$ & $0.20$  & $-$  & $-0.92$
\\
$\Lambda_{c}^{+}\to p K^{0}$ & $-0.13$  & $0.24$ & $0.11$ & $0.40$ & $-0.51$ & $-0.11$ & $0.04$  & $-$  & $-0.65$
\\
$\Lambda_{c}^{+}\to n K^{+}$ & $0.36$  & $-0.24$ & $0.12$ & $-1.13$ & $0.51$ & $-0.62$ & $0.21$  & $-$  & $-0.77$
\\
\bottomrule
%\hline\hline
\end{tabular}
%\end{ruledtabular}
%\end{center}
% }
\end{table}
%=========================================

Based on analytical equations Eq. (\ref{eq:Gamma}) and relevant expressions 
for each component, now we shall numerically calculate branching fractions and up-down decay asymmetries.
The decay asymmetries purely depend on $S$- and $P$-wave amplitudes while branching fractions rely on 
 more parameters, lifetimes. %The latest values have been given in Eq. (\ref{eq:lifetime}).
 In this work, the values of lifetime are taken as the new world averages 
 (in units of $10^{-13} \,\mathrm{s}$)
\begin{align} \label{eq:lifetime}
\tau(\Lambda_c^+)=2.03\pm0.02, \qquad
\tau(\Xi_c^+)=4.56\pm0.05, \qquad
\tau(\Xi_c^0)=1.53\pm0.02.
\end{align}
Especially note that the measured $\Xi_c^0$ lifetime by the LHCb is approximately 3.3 standard deviations larger than the old world average value \cite{Tanabashi:2018oca}.

All the channels with a kaon in the final states  receive both factorizable and nonfactorizable contributions.
For the factorizable amplitudes $A^{\rm{fac}}$ and $B^{\rm{fac}}$, their signs are 
co-determined by effective Wilson coefficients $a_1, a_2$ and FFs $f_1, g_1$.
%Let's first to di
The flipped sign between $\Sigma^0 K^+$ and $\Lambda^0 K^+$ is due to the sign difference 
between FFs, sharing common effective Wilson coefficient $a_1$. 
And for the two modes $\Sigma^+ K^0$ and $\Lambda^0 K^+$, their FFs  are both negative but differ from effective Wilson coefficients, hence their factorizable $S$-wave amplitudes 
are also with wrong sign. On the other side, the signs of nanfactorizable terms of the  three modes are the same for both $S$- and $P$-wave amplitudes. Thus with constructive
interference, the predictions of branching fractions for the two modes $\Sigma^+ K^0$ and $\Sigma^0 K^+$  is one order of magnitude larger than $\Lambda^0 K^+$, which receive a destructive interference. 
The situation is similar in the case of $\Xi_c^0$ decays.
However, for the two decaying modes of $\Lambda_c^+$, factorizable and nonfactorizable contributions 
to both $S$- and $P$-wave amplitudes have opposite sign thus a cancellation
occurs, leading to the smaller magnitudes of the branching ratios of $10^{-5}$ or even $10^{-6}$.
Among all the DCS decay modes, the three channels $\Xi_c^+\to \Sigma^+ K^0$, 
$\Xi_c^+\to\Sigma^0 K^+$ and $\Xi_c^0\to \Sigma^- K^+$ are predicted to be
most accessible by future experiments, as large as $10^{-4}$ in magnitude for their branching fractions.
The decay asymmetries, on the other hand, are all predicted to be negative in sign and 
larger than $0.5$ in magnitude.

%\vspace{2cm}
%===============

The contribution from $W$-exchange diagram $E_3$ in Fig. \ref{fig:anti-triplet} has been neglected throughout the whole calculation. This feature was first pointed out by K\"orner and Kr\"amer \cite{Korner:1992wi}
and argued by Zenczykowski \cite{Zenczykowski:1993jm} according to spin-flavor structure. 
A recent global fitting in
terms of topological diagrams also indicates the smallness of $E_3$ \cite{Zhao:2018mov}.
By dropping $E_3$ contribution, which induces  strong cancellations in both $S$- and $P$-wave amplitudes, the long-standing puzzle in $\Lambda_c^+\to \Xi^0 K^+$ has been successfully resolved recently \cite{Zou:2019kzq}. Hence in this work, we continue working in this scheme and find 
 the two modes $\Xi_c^+\to  n\pi^+$ and $\Xi_c^0\to p \pi^-$ are manifestly affected. 
 A straightforward calculation of commutators  in $S$-wave is subject to a strong cancellation
 between $a_{p\Xi_c^+}$ and $a_{n\Xi_c^0}$.
 Such a cancellation can be understood from the two pole diagrams corresponding to $E_3$.  
 Due to the aforementioned reasons, $E_3$ can be neglected and hence
 the cancellation can be avoided.
 The remaining results then are obtained by
% However, one should also 
taking into account another $W$-exchange diagram $E_1$ for 
 $\Xi_c^+\to  n\pi^+$  and
 $E_2$ for $\Xi_c^0\to  p\pi^-$. 
 %Due to the aforementioned reasons, such a cancellation
 %can be avoided by neglecting $E_3$ contributions.
 %the contribution from $E_3$ can be neglected.
 %
 % still contributes to $S$-wave amplitude.
%Hence  by ascribing to $E_3$  and dropping $E_3$ effectively, the cancellation 
 %can be avoided, %just like the treatment aforementioned,
  %
 % Likewise, due to the reason aforementioned,  such a cancellation can be avoided by neglecting $E_3$, 
 %leading to a not so small $S$-wave
 %amplitude. 
 Especially for the mode $\Xi_c^+\to n\pi^+$, its branching fractions is predicted to be close to $1\times 10^{-4}$ and decay asymmetries are large in magnitude and positive in sign, 
 which is possibly accessible by BESIII or Belle-II in the near future. 
 
%===============

The topological diagrams have revealed that the modes containing a pion or $\eta$ receive purely nonfactorizable contributions. Furthermore, the calculation in the soft-pseudoscalar limit
indicates that the nonfactorizable $S$-wave of all modes with neutral pseudoscalar vanish. 
%There are two reasons for such a consequence.
%For the channels with neutral meson final states, 
The identical quantum numbers (third-component of isospin or hypercharge) of initial and final baryons lead to  vanishing baryon matrix elements of commutators, which is a 
natural consequence of current algebra. 
This feature of current algebra calculation, however, is less reliable for the four modes.
At least a correction of current algebra result may induce an obvious different prediction
to decay asymmetry. An exact pole model estimation will be carried out in our future work.

%\subsection{Final states with $K_S$ and $K_L$}

From the experimental point of view, the measured neutral kaon is actually 
%more direct touched neutral kaon are 
$K_S$ with its lifetime
$\tau=8.954\times 10^{-11}$s.  From the relation between $K_{S,L}$
and $K^0, \overline{K}^0$
\begin{align}
& K_S= \frac{1}{\sqrt{2}}\left(\frac{1+\epsilon}{\sqrt{1+|\epsilon|^2}} K^0
+\frac{-1+\epsilon}{\sqrt{1+|\epsilon|^2}} \overline{K}^0\right),\nonumber\\
& K_L= \frac{1}{\sqrt{2}}\left(\frac{1+\epsilon}{\sqrt{1+|\epsilon|^2}} K^0
+\frac{1-\epsilon}{\sqrt{1+|\epsilon|^2}} \overline{K}^0\right),
\end{align}
together with  $|\epsilon|=(2.228\pm 0.011)\times 10^{-3}$ \cite{Tanabashi:2018oca}, 
we can get 
\begin{equation}
{\mathrm{Br}}(\mathcal{B}_c \to \mathcal{B} K_S) \approx
 \frac12 {\mathrm{Br}}(\mathcal{B}_c \to \mathcal{B} K^0) + \frac12{\mathrm{Br}}(\mathcal{B}_c \to \mathcal{B} \overline{K}^0).
 \end{equation}
For the $\Lambda_c^+$ decays, it is interesting to notice that 
$\Lambda_c\to p \overline{K}^0$ is CF process with
the predicted branching fraction $2.11\times 10^{-2}$ \cite{Zou:2019kzq} while
$\Lambda_c \to p K^0$ is DCS one with 
branching fraction $4\times 10^{-6}$ predicted in current work. 
This huge but natural difference between the two modes hence brings difficulty  to extract the data of 
$\Lambda_c^+\to p K^0$ from the measurement of $\Lambda_c^+ \to p K_S$ in BESIII.
The similar situation occurs in the decays $\Xi_c^0 \to\Lambda^0 \overline{K}^0$
and $\Xi_c^0 \to\Lambda^0 {K}^0$.
Fortunately, there are  two exceptions, 
\begin{align}
&{\mathrm{Br}}(\Xi_c^+ \to \Sigma^+ \overline{K}^0)= 2\times 10^{-3}\; 
({\mathrm{CF}}),\quad
{\mathrm{Br}}(\Xi_c^+ \to \Sigma^+ {K}^0)= 1\times 10^{-4}\; ({\mathrm{DCS}}),\nonumber\\
&{\mathrm{Br}}(\Xi_c^0 \to \Sigma^0 \overline{K}^0)= 4\times 10^{-4}\; 
({\mathrm{CF}}),\quad ~
{\mathrm{Br}}(\Xi_c^0 \to \Sigma^0 {K}^0)= 2\times 10^{-5}\; ({\mathrm{DCS}}),
\end{align}
%\begin{equation*}
%{\mathrm{Br}}(\Xi_c^+ \to \Sigma^+ \overline{K}^0)= 2\times 10^{-3}\; 
%({\mathrm{CF}}, \cite{Zou:2019kzq}),\quad
%{\mathrm{Br}}(\Xi_c^+ \to \Sigma^+ {K}^0)= 1\times 10^{-4}\; ({\mathrm{DCS}}),
%\end{equation*}
%\begin{equation}
%\hspace{0.52cm} {\mathrm{Br}}(\Xi_c^0 \to \Sigma^0 \overline{K}^0)= 4\times 10^{-4}\; 
%({\mathrm{CF}}, \cite{Zou:2019kzq}),\quad ~
%{\mathrm{Br}}(\Xi_c^0 \to \Sigma^0 {K}^0)= 2\times 10^{-5}\; ({\mathrm{DCS}}),
%\end{equation}
in which the differences between CF (see \cite{Zou:2019kzq}) and DCS modes are not dramatically huge.
Thus it is hopeful to measure the two DCS modes, especially $\Xi_c^+ \to \Sigma^+ K^0$,  when more data are accumulated.  To be specific, we can also give predictions 
\begin{equation}
{\mathrm{Br}}(\Xi_c^+ \to \Sigma^+ K_S)= 1.1\times 10^{-3},\quad
{\mathrm{Br}}(\Xi_c^0 \to \Sigma^0 K_S)= 2.1\times 10^{-4},
\end{equation}
which can be tested by Belle-II or BESIII in the near future.

%\newpage
\subsection{Comparison with other works}

%=========================

\begin{table}[h]
\begin{center}
%\footnotesize{
\caption{Comparison with other works for branching fractions in unit of $10^{-5}$ and decay asymmetries shown in parentheses.
}\label{tab:comparison}
 \vspace{0.3cm}
%\begin{ruledtabular}
\begin{tabular}{l *4{c}}% c c  }
\toprule
Modes& Our work  & Geng \emph{et al.}\cite{Geng:2019xbo} &Zhao \emph{et al.}\cite{Zhao:2018mov} &Experiment\\%
\midrule
%\hline
$\Xi_{c}^{+}\to p \eta$ & $3.2(0)$  & $19.8\pm7.6(-0.58\pm0.12)$  & $16.6\pm3.1$&$-$
\\
$\Xi_{c}^{+}\to p \pi^{0}$  & $1.2(0)$  & $5.3\pm1.2(0.81\pm0.12)$  & $1.5\pm1.5$&$-$
\\
$\Xi_{c}^{+}\to n \pi^{+}$  & $8.8(0.88)$  & $10.7\pm2.4(0.81\pm0.12)$  & $5.2\pm1.5$&$-$
\\
$\Xi_{c}^{+}\to\Sigma^{+}K^{0}$  & $12.8(-0.79)$  & $18.6\pm1.6(-0.96^{+0.11}_{-0.04})$  & $16.9\pm5.4$&$-$
\\
$\Xi_{c}^{+}\to\Sigma^{0}K^{+}$  & $22.6(-0.96)$  & $12.1\pm0.6(-1.00^{+0.02}_{-0.0})$  & $7.2\pm1.8$&$-$
\\
$\Xi_{c}^{+}\to\Lambda^{0}K^{+}$ & $1.8(-0.54)$  & $3.1\pm0.5(0.50\pm0.16)$  & $7.5\pm1.9$&$-$
\\
$\Xi_{c}^{0}\to n \eta$   & $1.1(0)$  & $6.6\pm2.5(-0.58\pm0.12)$  & $4.2\pm0.8$&$-$
\\
$\Xi_{c}^{0}\to n \pi^{0}$    & $0.4(0)$ &$1.8\pm0.4(0.81\pm0.12)$  & $3.3\pm0.9$&$-$
\\
$\Xi_{c}^{0}\to p \pi^{-}$   & $3.0(0.88)$  & $3.6\pm0.8(0.81\pm0.12)$  & $7.6\pm2.0$&$-$
\\
$\Xi_{c}^{0}\to\Lambda^{0}K^{0}$  & $2(-0.92)$  & $0.9\pm0.3(0.00\pm0.33)$  & $2.4\pm1.4$&$-$
\\
$\Xi_{c}^{0}\to\Sigma^{0}K^{0}$  & $2.2(-0.79)$  & $3.1\pm0.3(-0.96^{+0.11}_{-0.04})$  & $2.3\pm1.4$&$-$
\\
$\Xi_{c}^{0}\to\Sigma^{-}K^{+}$  & $15.2(-0.96)$  & $8.1\pm0.4(-1.00^{+0.02}_{-0})$  & $5.5\pm0.7$&$-$
\\
$\Lambda_{c}^{+}\to p K^{0}$  & $0.4(-0.65)$  & $0.8\pm1.1(0.97^{+0.03}_{-0.12})$  & $3.7\pm1.1$&$-$
\\
$\Lambda_{c}^{+}\to n K^{+}$  & $2.1(-0.77)$  & $0.5\pm0.2(-0.61^{+0.76}_{-0.39})$  & $1.4\pm0.5$&$-$
\\
\bottomrule
\end{tabular}
%\end{ruledtabular}
%}
\end{center}
\end{table}

%=========================

Weak decays of charmed baryons have attracted many interests in recent time.
Based on SU(3) flavor symmetry in theory and taking measured branching fractions and
asymmetries
as inputs, predictions of more branching fractions and decay asymmetries can be
obtained in a global fitting picture\cite{Geng:2019xbo}. 
A recent new exploration 
by parameterizing topological diagrams with independent parameters also
provides another set of predictions of branching fractions \cite{Zhao:2018mov}.  
In this part  a comparison with these groups is made
and shown in Table \ref{tab:comparison}.

There are some common features for the three groups relying on different approaches.
First the branching fractions of DCS decays are all ranged in $10^{-6} \sim 10^{-4}$.
Especially for the decays $\Xi_c^+\to \Sigma^+ K^0$, $\Xi_c^+\to \Sigma^0 K^+$ and
$\Xi_c^0\to \Sigma^- K^+$,
 all the three groups agree that
their branching fraction are of $(1\sim 2)\times 10^{-4}$ and with large and negative decay asymmetries, which is highly accessible by Belle-II or BESIII in the near future. 
For the two modes $\Xi_c^+ \to n\pi^+$ and $\Xi_c^0\to p \pi^-$, 
not only their branching fractions agree well for the three groups,
 but also the size as well as the sign of decay asymmetries can be confirmed
 by two independent groups.
 %
% {\color{blue}
 The consistent predictions for both branching ratios and decay asymmetries
 of $\Xi_c^+ \to n\pi^+$ and $\Xi_c^0\to p \pi^-$, on the other hand, 
 confirms our treatment by neglecting $E_3$.
% }
 %
The size of $\Lambda_c^+\to nK^+$,  however, are of $(1\sim 2)\times 10^{-5}$ in magnitude
and also with large negative asymmetry.  
It is known that BESIII can reconstruct a
neutron final state. With more data accumulated after its upgrade, this channel could possibly be measured \cite{Ablikim:2019hff}.
 
There are also some disagreement in the three modes
$\Xi_c^+\to \Lambda^0 K^+$, $\Xi_c^0\to \Lambda^0 K^0$ and $\Lambda_c^+\to p K^0$.
Our prediction for the branching fractions of $\Xi_c^+\to \Lambda^0 K^+$ and $\Lambda_c^+ \to p K^0$ are the smallest among all the three groups, while for $\Xi_c^0\to \Lambda^0 K^0$ ours
is close to the prediction in \cite{Zhao:2018mov}. For the decay asymmetries,
ours differ from \cite{Geng:2019xbo} for the signs in the two modes 
$\Xi_c^+\to \Lambda^0 K^+$ and $\Lambda_ c^+\to p K^0$,
and magnitude for the mode $\Xi_c^0\to \Lambda^0 K^0$.

%Our prediction of $\Xi_c^+\to p \eta$ is about 
%one order of magnitude smaller than the other two groups. Also the zero 
%decay asymmetries in the modes containing pion or $\eta$ have not been observed 
%in \cite{Geng:2019xbo}. The situation would be converged when more  data are available.

\section{Conclusions}
\label{sec:con}

In this work we have studied the branching fractions and up-down decay asymmetries of  DCS decays of antitriplet charmed baryons.
In the topological-diagram approach, we can identify factorizable and nonfactorizable 
contributions in each process clearly.
The calculation of factorizable and nonfactorizable terms in
$S$- and $P$-wave amplitudes is carried out in separated ways. 
For the factorizable amplitudes, by defining an effective color number $N_c$ encoded in effective Wilson coefficients, one 
can make use of naive factorization.
To estimate nonfactorizable contribution, we work in the pole model for $P$-wave amplitudes and current algebra
for $S$-wave ones. All the non-perturbative parameters, including baryon-baryon transition form factors, baryon matrix elements and axial-vector form factors, are evaluated within the MIT bag model throughout the whole calculations.

Some conclusions can be drawn from our analysis as follows.

\begin{itemize}

\item 
The decays $\Xi_c^+\to \Sigma^+ K^0, \Sigma^0 K^+$ and $\Xi_c^0\to \Sigma^- K^+$ are the most
promising DCS channels to be measured   as their branching fractions are predicted to be as large as 
$(1\sim 2) \times 10^{-4}$, which agree with the other predictions based on different approaches. 
The decay asymmetries are found to be large in magnitude and negative in sign. 

\item 
For the decay channels containing $K^0$ in the final states, 
it is possible to extract $\Xi_c^+\to \Sigma^+ K^0$ and $\Xi_c^0\to \Sigma^0 K^0$
from data with $K_S$ in the final states.
However, the measurement of $\Lambda_c^+\to p K^0$ and $\Xi_c^0\to \Lambda^0 K^0$
in experiment is challenging. 

\item
Predictions for the two modes $\Xi_c^+\to n\pi^+, \Xi_c^0 \to p \pi^-$ agree well 
among three different groups, both for their branching fractions and decay asymmetries.
Though with small but anticipated branching fraction of $10^{-5}$,
a further confirmation from experiment will be significant to clarify the dynamic mechanism
at the quark level.

\end{itemize}

\begin{acknowledgments}

We would like to thank Prof. Hai-Yang Cheng for his encouragement and fruitful discussion on this work.
This research  is supported by NSFC under Grant No. U1932104 and No. 11605076.

\end{acknowledgments}

%\newpage

\appendix

\section{Model estimation of non-perturbative parameters }
\label{app:nonp}
There are three types of non-perturbative quantities involved in charmed baryon decays:
the baryon transition form factors contributing to factorizable amplitudes and baryon matrix elements
as well as axial vector form factors contributing to non-factorizable amplitudes.
In this work, the estimation of these parameters are carried out
within the
framework of the MIT bag model \cite{MIT}.

\subsection{Baryon transition form factors}
\label{app:FF}

In the zero recoil limit where $q^2_{\rm{max}}=(m_i-m_f)^2$, FFs %can be
calculated in
%expressed within
the  MIT bag model \cite{Cheng:1993gf} are given as
\begin{align} \label{eq:f1g1}
& f_1^{\mathcal{B}_f \mathcal{B}_i}(q^2_{\rm{max}})=\la \mathcal{B}_f^\uparrow| b_{q_1}^\dagger b_{q_2}| \mathcal{B}_i^\uparrow\ra \int d^3 \bm{r}\Big(u_{q_1}(r)u_{q_2}(r)+v_{q_1}(r)v_{q_2}(r)\Big),
\nonumber\\
& g_1^{\mathcal{B}_f \mathcal{B}_i}(q^2_{\rm{max}})=\la \mathcal{B}_f^\uparrow| b_{q_1}^\dagger b_{q_2} \sigma_z| \mathcal{B}_i^\uparrow\ra \int d^3 \bm{r}
\left(u_{q_1}(r)u_{q_2}(r)-\frac13v_{q_1}(r)v_{q_2}(r)\right),
\end{align}
where $u(r)$ and $v(r)$ are the large and small components, respectively, of the quark wave function in the bag model.
The two quark flavors $q_1, q_2$ are determined by the meson content.
The physical FFs which contribute to the factorizable amplitudes are actually located at energy scale $q^2=m_P^2$, thus 
an evolution from different energy scale is necessary. Follow \cite{Cheng:1991sn},
the connections of FFs at different scale are
\begin{equation}
f_i(q^2)=\frac{f_i(0)}{(1-q^2/m_V^2)^2},\qquad
g_i(q^2)=\frac{g_i(0)}{(1-q^2/m_A^2)^2},
\label{eq:FF1}
\end{equation}
where $m_V%=m_{D_s^*}
=2.01\,{\rm GeV}$, $m_A%=m_{D_{s1}(2536)}
=2.42\,{\rm GeV}$ for the  $(c\bar{d})$ quark content, and
$m_V=2.11\,{\rm GeV}$, $m_A=2.54\,{\rm GeV}$ for $(c\bar{s})$ quark content.

\begin{table}[b]
\begin{center}
\linespread{1.5}
\footnotesize{
 \caption{The calculated form factors in the MIT bag model at maximum four-momentum
 transfer squared $q^2=q^2_{\rm{max}}=(m_i-m_f)^2$ and $q^2=m_P^2$.
 } \label{tab:FF}
\vspace{0.3cm}
%\begin{ruledtabular}
\begin{tabular}%{\textwidth}
{ cccc|ccc}
% \hline\hline
\toprule
 modes &  $f_1(q_{\rm{max}}^2)$ &$ f_1(m_P^2)/f_1(q_{\rm{max}}^2)$ & $f_1(m_P^2)$ &  $g_1(q_{\rm{max}}^2)$ &
 $g_1(m_P^2)/g_1(q_{\rm{max}}^2)$ & $g_1(m_P^2)$  \\
 %\hline
 \midrule
%\colrule
$\Xi_{c}^{+}\to\Sigma^{0}K^{+}$  & $\frac{\sqrt{3}}{2}Y_{1}$  &$0.404$ &$0.308$    & $\frac{\sqrt{3}}{2}Y_{2}$  &   $0.568$&$0.378$   \\
$\Xi_{c}^{+}\to\Lambda^{0}K^{+}$   & $-\frac{1}{2}Y_{1}$  &$0.338$  &$-0.150$  & $-\frac{1}{2}Y_{2}$  &$0.515$ &$-0.198$   \\
$\Xi_{c}^{+}\to\Sigma^{+}K^{0}$    & $-\frac{\sqrt{6}}{2}Y_{1}$  &$0.401$ &$-0.433$   & $-\frac{\sqrt{6}}{2}Y_{2}$  &$0.567$ &$-0.534$   \\
$\Xi_{c}^{0}\to\Sigma^{-}K^{+}$    & $\frac{\sqrt{6}}{2}Y_{1}$  &$0.404$ &$0.437$   & $\frac{\sqrt{6}}{2}Y_{2}$  &$0.570$&$0.536$   \\
$\Xi_{c}^{0}\to\Sigma^{0}K^{0}$   & $\frac{\sqrt{3}}{2}Y_{1}$  &$0.401$ &$0.306$   & $\frac{\sqrt{3}}{2}Y_{2}$  &$0.567$&$0.377$   \\
$\Xi_{c}^{0}\to\Lambda^{0}K^{0}$  & $\frac{1}{2}Y_{1}$  &$0.336$ &$0.148$    & $\frac{1}{2}Y_{2}$  &$0.514$&$0.198$   \\
$\Lambda_{c}^{+}\to p K^{0}$    & $-\frac{\sqrt{6}}{2}Y_{1}$  &$0.342$ &$-0.369$    & $-\frac{\sqrt{6}}{2}Y_{2}$  &$0.519$ &$-0.488$   \\
$\Lambda_{c}^{+}\to n K^{+}$    & $-\frac{\sqrt{6}}{2}Y_{1}$  &$0.342$ &$-0.370$   & $-\frac{\sqrt{6}}{2}Y_{2}$  &$0.519$ &$-0.489$   \\
\bottomrule
%\hline\hline
\end{tabular}
%\end{ruledtabular}
}
\end{center}
\end{table}

It is obvious that the FF at $q^2_{\rm{max}}$ is determined only by the baryons in initial and final states.
However, its evolution with $q^2$ is governed by both the final-state meson and relevant quark content. 
This feature manifests in
Table \ref{tab:FF}, where 
 the FFs calculated at $q^2_{\rm{max}}$ in the zero recoil limit %, according to Eq. (\ref{eq:FF1})
are presented in the second and fifth columns. 
The auxiliary quantities $Y_{1,2}$ can be obtained from the calculation in MIT bag model,  giving
\begin{equation}
Y_1=4\pi \int r^2 dr (u_u u_c+v_u v_c),\quad~~ 
Y_2=4\pi \int r^2 dr (u_u u_c-\frac13 v_u v_c).
\end{equation}
The model parameters are adopted from \cite{Cheng:2018hwl} and references therein. Numerically, we have $Y_1=0.88, Y_1^s=0.95,
Y_2=0.77
, Y_2^s=0.86
$, which are consistent with
the corresponding numbers in \cite{Cheng:1993gf}.
The evolution constants are shown in third and sixth columns. And  in fourth and seventh columns, we list  physical FFs. 

\subsection{Baryon matrix elements}
\label{app:bme}
The  baryonic matrix elements $a_{\B'\B}$ get involved  both in $S$-
and $P$-wave amplitudes.
Their general expression in terms of the effective Hamiltonian Eq. (\ref{eq:Hamiltonian}) is given by
\begin{equation}
a_{\B'\B} \equiv  \la \B'|\mathcal{H}_{\rm{eff}}^{\rm{PC}}|\B\ra
=\frac{G_F}{2\sqrt{2}} V_{cd} V^*_{us} c_-\la \B' |O_- |\B\ra,
\end{equation}
where  and
$O_\pm=(\bar{d}c)(\bar{u}s)\pm(\bar{d}s)(\bar{u}c)$
 and $c_\pm=c_1\pm c_2$.
The matrix element of $O_+$ vanishes as this operator is symmetric in color indices. 
The further calculation of relevant baryon matrix elements is carried out in MIT bag model
(see appendix of Ref. \cite{Cheng:2018hwl}),
and results are
\begin{align}
 & \langle p|O_{-}|\Xi_{c}^{+}\rangle =-2\sqrt{\frac{2}{3}}(X_{1}^{D}+3X_{2}^{D}),\hspace{1cm}
  \langle n|O_{-}|\Xi_{c}^{0}\rangle =2\sqrt{\frac{2}{3}}(X_{1}^D-3X_{2}^{D}),\nonumber\\
 & \langle p|O_{-}|\Xi_{c}^{'+}\rangle = -\frac{2}{3}\sqrt{2}(X_{1}^D-9X_{2}^D),\hspace{1cm}
 \langle n|O_{-}|\Xi^{'0}_{c}\rangle =\frac{2}{3}\sqrt{2}(X_{1}^D+9X_{2}^D),\nonumber\\
% & \langle \Lambda^{0}|O_{-}|\Xi^{-}\rangle =0 ,\hspace{4.25cm}
 %\langle \Sigma^{0}|O_{-}|\Xi^{-}\rangle =0 \nonumber\\
 & \langle \Sigma^{0}|O_{-}|\Omega_{c}^{0}\rangle =\frac{4}{3}\sqrt{2}X_{1}^D,\hspace{3cm}
\langle \Lambda^{0}|O_{-}|\Omega_{c}^{0}\rangle = -4\sqrt{6}X_{2}^D,
 \end{align}
where we have introduced the bag integrals $X_1^D$ and $X_2^D$ as
\begin{align} \label{eq:X}
&X_1^D= \int^R_0r^2 dr (u_uv_u-v_u u_u)(u_c v_s -v_c u_s),\nonumber\\
&X_2^D= \int^R_0r^2 dr (u_u u_u+v_u v_u)(u_c u_s +v_c v_s),
\end{align}
with the numbers  $X_1^D=0, X_2^D= 1.78\times 10^{-4}$.
To obtain numerical results, we have employed the following bag parameters
\be
m_u=m_d=0, \quad m_s=0.279~{\rm GeV}, \quad m_c=1.551~{\rm GeV}, \quad R=5~{\rm GeV}^{-1},
\en
where $R$ is the radius of the bag.

\subsection{Axial-vector form factors}
\label{app:avff}
Here we directly show the results  of MIT bag model estimation of axial-vector form factors,
\begin{align}
-2g^{A(\eta_{8})}_{\Xi_{c}^{'+}\Xi_{c}^{+}}& = \frac{6}{5}g^{A(\pi^{0})}_{PP}=-2\sqrt{3}g^{A(\pi^{0})}_{\Xi_c^{'+} \Xi_c^+}
=\frac{3}{5}g^{A(\pi^{+})}_{np}=-\sqrt{3}g^{A(\pi^+)}_{\Xi_{c}^{'0}\Xi_{c}^{+}}
=-2g^{A(\eta_{8})}_{\Xi_{c}^{'0}\Xi_{c}^{0}}\nonumber\\ 
&=-\frac{6}{5}g^{A(\pi^{0})}_{nn}=2\sqrt{3}g^{A(\pi^{0})}_{\Xi_c^{'0} \Xi_c^0}=\frac{3}{5}g^{A(\pi^{-})}_{Pn}=-\sqrt{3}g^{A(\pi^-)}_{\Xi_{c}^{'+}\Xi_{c}^{0}}=2\sqrt{3}g^{A(\eta_{8})}_{PP}
\nonumber\\
&=2\sqrt{3}g^{A(\eta_{8})}_{nn}=(4\pi)Z_1, \\ 
 3\sqrt{2}g^{A(K^{+})}_{\Sigma^{0}P}& = -\frac{\sqrt{6}}{3}g^{A(K^{+})}_{\Lambda^{0} P}=
3g^{A(K^{0})}_{\Sigma^{+} P}=3g^{A(K^{+})}_{\Sigma^{-} n}=-\frac{\sqrt{6}}{2}g^{A(K^{0})}_{\Omega_{c}^{0} \Xi_{c}^{0}}\nonumber\\
&=-\frac{\sqrt{6}}{3}g^{A(K^{0})}_{\Lambda^{0} n} 
=-\sqrt{3}g^{A(K^{0})}_{\Xi^{'+}_{c} \Lambda_{c}^{+}}=
-\frac{\sqrt{6}}{2}g^{A(K^{+})}_{\Omega^{0}_{c} \Xi_{c}^{+}}
=-3\sqrt{2}g^{A(K^{0})}_{\Sigma^{0} n}\nonumber\\
&=\sqrt{3}g^{A(K^{+})}_{\Xi^{'0}_{c} \Lambda_{c}^{+}}=(4\pi)Z_2,\nonumber\\ 
g^{A(\eta_{8})}_{\Xi_{c}^{+}\Xi_{c}^{+}}&=g^{A(\pi^{0})}_{\Xi_c^+ \Xi_c^+}=g^{A(\pi^{+})}_{\Xi_c^{0} \Xi_c^+}=g^{A(K^{+})}_{\Xi^{-} \Xi_{c}^{+}}
=g^{A(\eta_{8})}_{\Xi_{c}^{0}\Xi_{c}^{0}}=g^{A(\pi^{0})}_{\Xi_{c}^{0}\Xi_{c}^{0}}
=g^{A(K^{+})}_{\Xi^{0}_{c} \Lambda_{c}^{+}}
%\nonumber\\
%&
=g^{A(\pi^{-})}_{\Xi_c^+\Xi_c^0}=g^{A(K^{0})}_{\Xi^{+}_{c} \Lambda_{c}^{+}}=0.
\nonumber
\end{align}
where the auxiliary bag integrals are given by
\begin{equation}
Z_1=\int r^2 dr\left(u_u^2 -\frac13 v_u^2\right),\qquad
Z_2=\int r^2 dr \left(u_u u_s -\frac13 v_u v_s\right).
\end{equation}
Numerically,
$(4\pi) Z_1= 0.65$ and  $(4\pi)Z_2=0.71$.
Our results in the last equation also confirm
the vanishing coupling between antritriplet baryons. 
%a general features $g_{\bar{3}\bar{3}}^P=0$
%\newpage

\end{document}